\documentstyle[epsf]{mn}
\newif\ifAMStwofonts
\ifoldfss

  \ifCUPmtlplainloaded \else
    \NewTextAlphabet{textbfit} {cmbxti10} {}
    \NewTextAlphabet{textbfss} {cmssbx10} {}
    \NewMathAlphabet{mathbfit} {cmbxti10} {} % for math mode
    \NewMathAlphabet{mathbfss} {cmssbx10} {} %  "   "    "
  \fi
  \ifAMStwofonts
    \ifCUPmtlplainloaded \else
      \NewSymbolFont{upmath} {eurm10}
      \NewSymbolFont{AMSa} {msam10}
      \NewMathSymbol{\upi}     {0}{upmath}{19}
      \NewMathSymbol{\umu}     {0}{upmath}{16}
      \NewMathSymbol{\upartial}{0}{upmath}{40}
      \NewMathSymbol{\leqslant}{3}{AMSa}{36}
      \NewMathSymbol{\geqslant}{3}{AMSa}{3E}

      \let\leq=\leqslant 
       
    \fi
  \fi
\fi % End of OFSS

\ifnfssone
  \newmathalphabet{\mathit}
  \addtoversion{normal}{\mathit}{cmr}{m}{it}
  \addtoversion{bold}{\mathit}{cmr}{bx}{it}
  \newmathalphabet{\mathbfit} % math mode version of \textbfit{..}
  \addtoversion{normal}{\mathbfit}{cmr}{bx}{it}
  \addtoversion{bold}{\mathbfit}{cmr}{bx}{it}
  \newmathalphabet{\mathbfss} % math mode version of \textbfss{..}
  \addtoversion{normal}{\mathbfss}{cmss}{bx}{n}
  \addtoversion{bold}{\mathbfss}{cmss}{bx}{n}
  \ifAMStwofonts
    \ifCUPmtlplainloaded \else
      \UseAMStwoboldmath
      \makeatletter
      \new@mathgroup\upmath@group
      \define@mathgroup\mv@normal\upmath@group{eur}{m}{n}
      \define@mathgroup\mv@bold\upmath@group{eur}{b}{n}
      \edef\UPM{\hexnumber\upmath@group}
      \new@mathgroup\amsa@group
      \define@mathgroup\mv@normal\amsa@group{msa}{m}{n}
      \define@mathgroup\mv@bold\amsa@group{msa}{m}{n}
      \edef\AMSa{\hexnumber\amsa@group}
      \makeatother
      \mathchardef\upi="0\UPM19
      \mathchardef\umu="0\UPM16
      \mathchardef\upartial="0\UPM40
      \mathchardef\leqslant="3\AMSa36
      \mathchardef\geqslant="3\AMSa3E

      \let\leq=\leqslant 

    \fi
  \fi
\fi % End of NFSS release 1

\ifnfsstwo
  \DeclareMathAlphabet{\mathbfit}{OT1}{cmr}{bx}{it}
  \SetMathAlphabet\mathbfit{bold}{OT1}{cmr}{bx}{it}
  \DeclareMathAlphabet{\mathbfss}{OT1}{cmss}{bx}{n}
  \SetMathAlphabet\mathbfss{bold}{OT1}{cmss}{bx}{n}
  \ifAMStwofonts
    \ifCUPmtlplainloaded \else
      \DeclareSymbolFont{UPM}{U}{eur}{m}{n}
      \SetSymbolFont{UPM}{bold}{U}{eur}{b}{n}
      \DeclareSymbolFont{AMSa}{U}{msa}{m}{n}
      \DeclareMathSymbol{\upi}{0}{UPM}{"19}
      \DeclareMathSymbol{\umu}{0}{UPM}{"16}
      \DeclareMathSymbol{\upartial}{0}{UPM}{"40}
      \DeclareMathSymbol{\leqslant}{3}{AMSa}{"36}
      \DeclareMathSymbol{\geqslant}{3}{AMSa}{"3E}

      \let\leq=\leqslant 

    \fi
  \fi
\fi % End of NFSS release 2
\ifCUPmtlplainloaded \else
  \ifAMStwofonts \else % If no AMS fonts
    \def\upi{\pi}
    \def\umu{\mu}
    \def\upartial{\partial}
  \fi
\fi

\title{Principal component analysis of {\it IUE} galaxy spectra}

\author [L.Formiggini and N. Brosch]
{Liliana Formiggini and  Noah Brosch
\\ The Wise Observatory and the School of Physics and Astronomy 
\\ Raymond and Beverly Sackler Faculty of Exact Sciences 
\\Tel Aviv University, Tel Aviv 69978, Israel}
\date{Accepted 2003 MMM DD,
 Received 2003 MMM DD,
      in original form 2003 MMM DD }

\pagerange{\pageref{firstpage}--\pageref{lastpage}}
\pubyear{2002}

\begin{document}

\def\etal{{\it et al.\ }}
\def\kms{$\rm km\, s^{-1}$}
\def\msol{M$_{\odot}$ }
\def\eg{{\it e.g.,}}
\def\ie{{\it i.e.,}}
\def\Halpha{H$\alpha$}
\def\Hbeta{H$\beta$}
\def\Hgamma{H$\gamma$}
\def\Hdelta{H$\delta$}
\def\Lya{Ly$\alpha$}
\def\Lyb{Ly$\beta$}
\def\lam{$\lambda$}

\maketitle
\label{firstpage}

\begin{abstract}
We  analyse the UV spectral energy distribution of a sample of normal
galaxies  listed in  the {\it IUE} INES Guide No. 2-Normal Galaxies
(Formiggini \& Brosch, 2000) using a Principal Component Analysis. 

The sample consists of the {\it IUE}-SW spectra of the central regions 
of  118 galaxies, where the {\it IUE} aperture included more than 1 per cent of the 
galaxy size. The principal components are associated with the main
components observed in the UV spectra of galaxies.
The first component, accounting for the largest source of diversity, can be 
associated with the UV  continuum emission. The second component represents 
the UV contribution  of an underlying evolved  stellar population.
The third component is sensitive to the amount of activity 
in the central regions of galaxies and measures the strength of star formation 
events.

In all the samples analysed here the principal component 
representative of star-forming activity accounts for a  significant percentage 
of the variance. The  fractional contribution to the SED by the evolved stars 
and by the young population are similar.

Projecting the SEDs onto their eigenspectra, we find 
that none of the coefficients of the principal components  can outline 
an internal correlation  or can correlate with the optical morphological types. 
In a sub-sample of 43 galaxies, consisting of almost only  compact 
and BCD galaxies, the third principal component defines a sequence related to the degree 
of starburst activity of the galaxy. 
\end{abstract}

\section {Introduction}

The history of UV astronomy shows that (a) only a few missions operated
and yielded significant data, and (b) most of these missions dealt with imaging
or photometry, with very limited spectral science. Among the latter, the {\it IUE} 
project stands out with more than 18 years of successful operation and with 
a yield of more than 10$^5$ spectra collected and uniformly analyzed. 
These spectra have been available on-line in the {\it IUE} final archives
and through the various astronomy data portals. The compilations from the
{\it IUE} data base offer a somewhat biased view of the UV spectral characteristics 
of astronomical objects.  
In particular, the INES Guide No. 2 ``Normal Galaxies'' (Formiggini \& Brosch
2000) offers a view of the UV spectral characteristics of the non-active galaxies.

The advent of the {\it GALEX} all-sky UV survey, with a strong component of
spectral science (Bianchi et al. 1999), argues for the development of new tools to 
understand the spectral energy distribution of galaxies in the UV. 
While traditionally UV spectra have hardly been used for classification, partly 
because of the lack of a large number of spectra of different objects, it is possible 
that the $\sim10^5$ galaxy spectra expected from {\it GALEX} would provide
the necessary data base to develop an independent and uniform classification
network using only UV information.

The question we posed ourselves was whether it is possible to predict the UV
spectrum of a galaxy by using exclusively information from the optical domain.
The question is relevant in the context of UV surveys, when one needs to estimate
the spectral detections, and also when one deals with huge data bases derived from 
a survey of a large fraction of the sky.

In the optical domain, the Hubble  diagram provides a satisfactory classification 
for local galaxies, with redshift $\leq2$, based on their appearance. 
This morphological  classification  however,  is a qualitative and somehow 
subjective analysis of the observable features of a galaxy. 
Classification of several hundreds of galaxies, performed by different 
experts (Lahav et al. 1995), results in a dispersion of more 
than one morphological type. For irregular and peculiar galaxies the  
situation is worse, because the Hubble classification does not define if 
an irregularity or  peculiarity is related to  shape,  such as 
asymmetry,  or to  surface brightness, such as the lack of a central
concentration.

Morphological classification depends also  on the photometric
filter. The appearance of galaxies in the UV is different from that
in the optical, and details emphasized  in the UV
often do  not agree with the Hubble classification.
For instance, UV images of spiral galaxies show detailed features in the 
spiral arms and in the bulge, such as knots of circumnuclear star-formation
regions, that cannot be clearly detected in the optical band.
Low redshift  UV  galaxies are  important  both for understanding
the local Universe and for the extrapolation to high redshift samples,

However, the higher percentage of irregular and merger  types found 
in high redshift  samples  (Brinchmann et al. 1998) indicates that 
the Hubble sequence is not sufficient to describe the complex UV 
morphology. Kuchinski et al (2000) analyzed a sample of 34 
nearby galaxies imaged with  UIT, and found that the far-UV morphology does 
not agree with the bins of the Hubble sequence.  The introduction of two 
additional parameters, to quantify  the asymmetry and the central concentration 
of galaxies, is suggested by Kuchinski et al (2001).

Classification can be performed also using the spectral energy distribution (SED). 
Galaxies of the same morphological type tend 
to have  similar stellar populations and this results in similar 
spectral features in the optical domain. Hence, a  spectral classification
based on the SED is related to the  physical properties of a galaxy,  such as
its stellar and  gas content.
An objective method for spectral classification is the  Principal 
Component Analysis (PCA), extensively applied to samples 
of templates and to galaxy models  . This technique was
able to retrieve regularities present in the optical spectra and to
define a spectral sequence for normal galaxies (Connolly et al. 1995;
Sodr\'{e} \& Cuevas 1997; Galaz \& de Lapparent 1998; Ronen,
 Aragon-Salamanca \& Lahav 1999).

In this paper we analyse  a set of UV  spectra  of galaxies 
from the {\it IUE}  final archive. This archive represents the most extended 
UV database on galaxies until new systematic UV surveys from 
space, such as that by {\it GALEX}, will be available. We use the 
INES Guide No.2-Normal Galaxies (Formiggini \& Brosch, 2000)\footnote{
Retrievable  electronically at 
http://wise-iue.tau.ac.il/$\sim$lili/Fnet/Fnet.html}.

A large effort has been expended in studying the {\it IUE} spectra of galaxies by 
Bonatto, Bica \& Alloin (1995) and Bonatto et al. (1996, 1998, 1999), by 
grouping the galaxies presenting 
spectral and optical morphological similarities and coadding the spectra in 
order to produce high signal-to-noise template spectra for subsets in each group.

In this work we  analyse a set of  direct  spectra of galaxies, without 
attempting to  build a-priori  templates. The only selection criterion adopted
is the coverage parameter, defined in section 2. This parameter evaluates the
area of the galaxy included in  the {\it IUE} spectrograph aperture.
The PCA method is described in section 3.
In section 4  we  discuss the  PCA results, and
the effect of  normalization on the sample and on  a subset with
good signal-to-noise. A data set with spectra covering a higher fraction
of the galaxy is
also analysed, and the Hubble optical morphological sequence is
compared to the UV spectral sequence in section 5.

\section{The  {\it IUE} sample}

  A suitable set of UV SEDs of galaxies covering a wide range of Hubble types
can be extracted from the {\it IUE} final archive. The data sets provided by 
INES  ({\it IUE} Newly Extracted Spectra) consist of low-resolution spectra 
extracted with an improved method from the line-by-line images of the 
{\it IUE} Final Archive, and of high-resolution spectra resampled to the 
low-resolution wavelength step.  A collection of UV spectra of 274 normal 
galaxies has been  compiled as the INES Guide No. 2 (Formiggini \& Brosch 2000). 
In this guide,  a representative UV spectrum for each galaxy
has been selected, combining the longest short-wavelength (SW) and
long-wavelength (LW) exposures of the galaxy, both taken trough the large
aperture.
The aperture position of the {\it IUE} apertures for the SW and LW spectra, that
were obtained independently and required repositioning of the satellite,
were checked with  aperture overlays  on the galaxy images. This 
procedure assured that  the entrance apertures of both SW and LW spectra
were centered on  the galaxy optical position or,  alternatively, refer to 
same physical region of the galaxy, such as the HII regions of NGC4449 and 
NGC5236.

We found some cases where the {\it IUE} aperture coordinates did not
correspond to the coordinates of the galaxy, and  the {\it IUE} did not,  in fact,
observe the object. Such is, for instance, the {\it IUE} spectrum of NGC 3077, where 
the  misplaced  aperture contains  a foreground star instead of the galaxy.

The large  apertures of the {\it IUE} spectrograph are 
10"$\times$20" ovals, each corresponding in area to a  circular
diaphragm with a diameter of 15.1 arcsec (Longo \& Capaccioli 1992),
and many of the galaxies observed are extended with respect to these large
apertures.
In order to estimate the fraction of the galaxy area observed by {\it IUE},
we calculated for each galaxy the ``coverage parameter'' C, defined  as the 
logarithmic  ratio between the surface area of the galaxy and the area of the
large {\it IUE} aperture. 

\begin{equation}
     C=log[\pi \times (D_{25}^{2} \times R_{25}) / (15.1)^{2} \pi/4] 
\end{equation}

Here D$_{25}$ and R$_{25}$ are the major axis and the axial ratio of the 
optical image of the galaxy as listed in NED, in units of arcsec. 
The numerator approximates the surface area of the galaxy, represented as 
an ellipse with the major and minor axes of the galaxy.

A C value of zero implies that the entire galaxy was measured by  the {\it IUE} 
spectrum. For faint galaxies, where the axes are not measurable, the C 
parameter is $\leq 0$. The histogram of the number of galaxies as a function 
of the C parameter (see Figure 5 of Formiggini \& Brosch, 2000) shows that,  
for 90 per cent of the sample of  274 normal galaxies in the INES Guide No.2, 
the {\it IUE} aperture covered less than  10 per cent of the galaxy.

For the present investigation, only the galaxies with a C value up to 2,
corresponding to a coverage of more than 1 per cent of the 
galaxy  were assembled from the sample of Formiggini \& Brosch (2000). 
This means that the spectra are dominated by the content of 
the very center of the light distribution of the galaxy.
A few cases  where the aperture position was miscentered were rejected. 
The total number of galaxies included in the sample analyzed here is 118. 
Table 1 lists the relevant data for the set.
Column two indicates to which sub-sample the specific galaxy belongs, where 1
represents membership in the sub-sample of relatively high S/N and 2 
represents membership 
in the sub-sample of high coverage parameter (see  section 4).
This sample, resulting from many different research projects, contains 
galaxies of all  morphological types, although it is obviously biased 
toward the brightest UV galaxies.
It is a representative sample  of UV galaxy spectra although it is not uniform,
 and the percentage of galaxies in each morphological class is not representative
of the galaxy population at large. 
Consequently, this sample is not suitable for luminosity function and density 
investigation.
A sub-sample of galaxies with C values up to 1, corresponding to a coverage of
more than 10 \% of the galaxy area, was selected for comparison. This
reduced sample  of 43 galaxies, consisting  mostly of BCD and compact objects,
is analyzed in section 4.2.

\begin{figure}
\begin{minipage}{80mm}
%fig1.xvgr
\centerline{\epsfxsize=2.5in\epsfbox{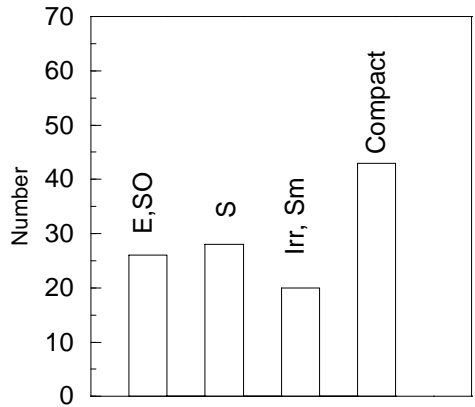}}
\caption{The distribution of spectral types}
\end{minipage}
\end{figure}

Figure 1  shows the morphological distribution of the data set. 
The galaxies have been binned in four morphological bins: ellipticals, 
spirals, irregulars, and a group containing all the compact objects,  
namely galaxies classified as BCD, EmLS, HII  and compact.
Note that the morphological types  and redshifts  adopted here are from 
NED, while in the INES Guide  No. 2 all information is from LEDA.
The sample includes all classical Hubble types, from elliptical to irregulars,
and  a large proportion of compact/BCD/EmLS galaxies whose emission in
the UV band is dominated by young, UV bright stars.

\begin{figure}
\begin{minipage}{80mm}
% isto_z.xvgr
\centerline{\epsfxsize=2.5in\epsfbox{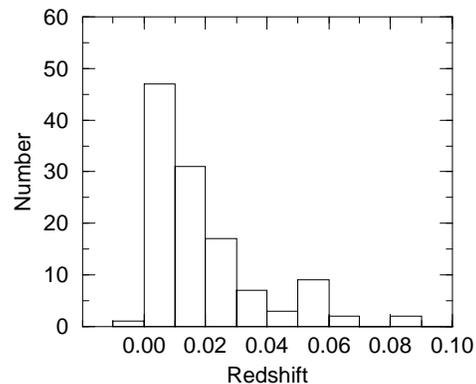}}
\caption{Histogram of the redshifts distribution}
\end{minipage}
\end{figure}

The redshift distribution  of the sample is shown in Figure 2.
{\it IUE} observed galaxies only in the very nearby Universe,
hence the sample analyzed in this paper can be considered as 
representing the local UV galaxy population. This sample can be used to
directly analyze the rest-frame UV SED of galaxies without requiring  
extrapolation from the optical data.

It is known that the galaxy spectra have a low signal-to-noise ratio
in the {\it IUE} LW region. Actually, a visual inspection of 
the spectra in the LW range, shows that the noise is the dominant 
feature. Furthermore, only part of the galaxy sample has been observed 
in the LW range.  Hence, we have restricted the analysis to the far-UV 
region covered by the  SW {\it IUE} images (between 1150--1900 \AA). 
This is the wavelength range where the most significant features probing
the galaxy evolution occur.

\section{The PCA method}
The technique used to analyze the UV spectra of our sample is the 
Principal Component Analysis (PCA; Whitney 1983).
The PCA algorithm is an orthogonal transformation that,  given a set 
of measured quantities, finds a set of new independent variables, called 
the principal components (PC), each one being a linear combination of the 
original quantities.  PCA has been demonstrated to be a useful tool for 
spectral classification of both stars (Ibata \& Irwin 1997; 
Singh, Gulati \& Gupta  1998) 
and galaxies (Connoly et al. 1995; Ronen et al. 1999).
It has been applied to quantities measured from the spectra, such as 
broad-band colors or equivalent width of lines (Turler \& Courvoisier 1998; 
Steindling, Brosch \& Rakos 2001), and also directly to the spectral energy distribution 
(SED) of galaxies (Sodr\'{e} \& Cuevas 1997; Galaz \& de Lapparent 1998).

In spectral PCA, each spectrum is  a vector in a multidimensional space,
and the fluxes in the individual wavelength bins represent the axes of this space. 
PCA identifies the  directions of the maximum variance  of the spectral 
components. These directions, called  principal components,  are identified 
sequentially and represent the new axes that  parameterize  the volume 
spanned  by the dataset. 
The principal components are ordered according to the fractional sample 
variance for which they account, the first component accounting for 
the largest fraction.
If there are common characteristics between the galaxy spectra of the sample, 
a few principal  components may account for a large fraction of the total 
sample variance. Therefore, the  PCA, by reducing the multi-dimensionality of 
the data, may provide a simple description of the original dataset and, through this, a
better insight on its underlying characteristics.

Each galaxy in our sample of {\it IUE} spectra is represented by a vector whose 
components are the  M  integrated fluxes in the individual wavelength 
channels. We consider a set of N vectors each of M elements 
[$ X_{\lambda,i}$, i=(1...,N), $\lambda$=(1...,M)],  where N is the 
number of galaxy spectra.
The sample is therefore represented by an N$\times$M matrix.
The eigenvectors  of the covariance matrix are the new axes, called the 
principal components (PC), and the eigenvalues give their variance.

Since for PCA the spectra must be identically  sampled in wavelength,  and 
should have an equal number of bins, the {\it IUE} data have been preprocessed. 
Each  spectrum has been dereddened for Galactic extinction using  
Seaton's (1979) law  
and shifted to the rest frame, according to the redshift data in Table 1.
The SEDs have been corrected only for interstellar extinction,
adopting  a conservative approach. It is well-known that 
the Seaton standard extinction law  is not suitable for
external galaxies, and  it sometimes overestimates the 2175~\AA~
bump (Calzetti, Kinney \& Storchi-Bergmann 1994, Calzetti 1999, 
Calzetti 2001). 
Moreover, the attenuation due to  dust either 
within the galaxy or along its line of sight is an important 
effect in the UV range. This attenuation effect depends 
strongly both on the relative geometric distribution of dust and 
stars, and on the dust chemical composition or metallicity. 
A differential internal reddening law should be applied to
each galaxy. However, we still have only a limited 
knowledge of the dust properties in the Milky Way and in 
the Magellanic Clouds, and of the differential extinction curve 
properties of quiescent and active-region galaxies (Gordon et al. 2003).
Since the  energy absorbed by dust is reprocessed, 
mainly in the infrared, an estimation of  the extinction requires 
multiwavelength data or, at least, a measure of the infrared 
emission for each galaxy. In the future such data will become 
available thanks to the UV and IR surveys of GALEX and of SIRTF.

Our sample of galaxies is  UV-selected, and we estimate that this 
ensures that the galaxy light is not heavily absorbed, 
although the presence of several dusty systems cannot be excluded.
 
Moreover, the ratio of the infrared  to the  far-UV emissions
is different for different star populations. 
%Reddening by dust has a degenerative effect since
A dust-reddened population mimics an older unreddened stellar 
population: this could affect the interpretation of the relative contribution
of each age population to the total galaxy spectrum.
Bonatto et al. (1998, 1999) already noted that some galaxies
have a contribution of young stars, although their spectrum 
looks very red. Spectral features, such as the line equivalent widths,
can help  distinguish  a reddened population from an old one.

The use of a statistical approach, such as the PCA, limits
the possibility of separating an older population from a reddened
one and this should be considered when analysing the possible
correlation (section 5) between  principal components and the 
morphological galaxy types. We believe that although the mixing
of morphological types can be partially attributed to differential
dust absorption,  our analysis indicates that the PCA methods
applied to larger galaxy samples could  be a useful classification 
method.

The redshift range spanned by the sample is small (see Fig 2), so that the wavelength 
range covered by all the galaxies of the sample is similar but slightly shorter than 
that covered by the SW {\it IUE} spectra.  Note that in almost
all galaxies, owing to the small redshift range, the NV \lam1242\AA~ line is blended
with the geocoronal Ly$\alpha$.  

The rest-frame spectra have been rebinned to a linear wavelength scale in 
the rest frame. The wavelength increment adopted is twice the sampling  interval of 
the low-resolution SW spectrograph of {\it IUE} ($\simeq$  3.4\AA), in order 
to preserve the spectral features of each spectrum. 
This increment determines the number of wavelength points to be used for 
the PCA. Note that the {\it IUE} spectral resolution is much superior to that of 
{\it GALEX} ($\Delta$$\lambda$/$\lambda$ $\simeq$150), Bianchi and 
the {\it GALEX} team, 1999) offering, in principle, a better discriminating power.

The sample chosen for the present analysis is therefore represented by 118 
galaxy spectra, each with 204 wavelength flux bins. 
The N$\times$M matrix representative of the sample is a 118$\times$204 singular matrix,
and  the eigenvalues of the covariance matrix were found using the 
singular valued decomposition technique (SDV:Mittaz, Penston \& Snijders 1990).
The average of the rest-frame, rebinned spectra has been subtracted from 
each spectrum. This procedure ensures that 
the dataset of the subtracted spectra should contain the maximal amount of 
discriminatory information. Otherwise, the average of the spectra dominates the 
matrix and the first principal component.

The problem of scaling and normalizing the input data before applying the PCA
method has been extensively addressed (Francis et al. 1992, Connoly et al. 1995).
For instance, the flux in each wavelength bin can be scaled to unit variance. 
The variance scaling is important when there are variables with a large range of 
absolute values, such as strong emission lines. The wavelength bins corresponding to 
these lines could dominate the PCA analysis and mask the continuum variations.
Francis et al. (1992) found that the PCA results are insensitive to scaling
and that the same number of components are present, although with different
fractions of the total variance.

Connoly et al. (1995) analysed the effect of various normalization  methods
on the results of the PCA using a sample of template galaxy spectra. The most popular
methods are the normalization by the integrated flux, or  by the  unit scalar product.
The normalization by the integrated flux is not suitable for our sample, since it 
would give the same weight to high signal-to-noise as to low signal-to-noise 
spectra. 

The alternative normalization method, 
by the scalar product, means that the sum of the squares of the (positive and
negative) fluxes across the spectrum is forced to unity,
 $\sqrt {\sum X_{\lambda,i}^{2}} = 1$

Fig. 3 shows the mean spectrum of the 118 galaxies sample before and after 
applying the scalar product normalization. While the continuum trend
is almost unchanged, the spectral  features are smeared out by this
 normalization procedure.

Since our data are flux-calibrated spectra, and the total flux is an intrinsic
characteristic of the galaxy, we applied the PCA analysis to the sample with 
no  normalization and no scaling  of the data. We compared the
results with that of a normalised sample and of a sub-sample of  galaxy spectra
with good signal-to-noise, normalised by the scalar product.

\begin{figure}
\begin{minipage}{80mm}
% fig 3
\centerline{\epsfxsize=3.0in\epsfbox{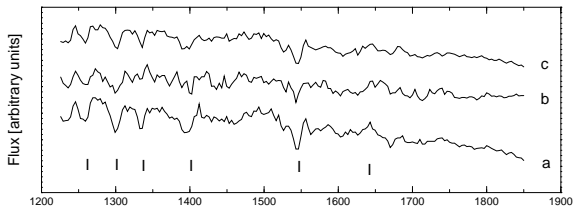}}
\caption{The mean spectrum of the sample without normalization (tracing a), 
after normalizing for the scalar product (tracing b), and after selecting 
for good signal-to-noise (tracing c). Tick marks indicates the following features:
SiI \lam1260, OI \lam1302 plus SiII \lam1304, CII \lam1335, SiIV \lam1400, 
CIV \lam1550, HeII \lam1640}
\end{minipage}
\end{figure}

\section { The principal components of the sample }

In this section, the input data for our analysis are the
fluxes in different wavelength bins, without
variance scaling or flux normalization. The principal components
are intrinsic to the spectra and can be related to both the magnitude and the
features of the observed spectra (Folkes, Lahav \& Maddox 1996).

\begin{figure}

% n_all_bin2.xvgr fig4
\centerline{\epsfxsize=3.5in\epsfbox{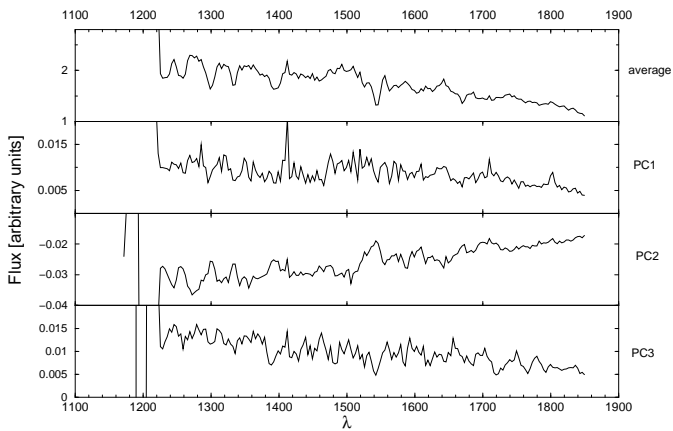}}
\caption{The average spectrum (top panel) and the principal components for the 
entire set. The three lower panels show the three principal components PC1, PC2, Pc3.}
%\end{minipage}
\end{figure}

The average spectrum (Fig. 4, top panel) shows no intense emission lines but only  
low and high ionization absorption lines, that can be both of stellar and
interstellar origin.
The most  prominent absorption lines are  the  SiIV \lam1400 and the  
CIV \lam1550 doublets, which show a complex structure reminiscent
of P-Cygni profiles.
Additional low-ionization
absorption  lines of the blend of  OI \lam1302 with SiII \lam1304,
 and of  CII \lam1335,  of both stellar
and interstellar origins, are  evident bluewards of 1400\AA~(De Mello, Leitherer \& Heckman 2000).
The emission feature at 1412 \AA ~ does not seem to be a spurious feature introduced
by noise. It could be the blend of SiIV \lam1400 with of SiIII \lam1417, 
a stellar photospheric line present in high-metallicity galaxies (Heckman et al. 1998).

The first three eigenvectors, shown  in Fig. 4 together with the
mean spectrum, account for 52 per cent, 17 per cent, and 14 per cent, respectively.  
These three  eigenvalues carry  83 per cent  of the information 
content  of the total variance.  Adding the fourth component, we account for  
up to 91 per cent of the variance. Subsequent eigenspectra  contribute
each a few per cent of the total variance.

The first PC (Fig. 4b) is flatter than the mean spectrum and modulations are present, 
such as the  emission feature at \lam1412  as well as some absorption features such as the 
SiII/OI \lam1300 and the  CII \lam1335 lines (Heckman et al. 1998). 
This PC component, accounting for the largest variance of the data, 
represents the diversity  from the average spectrum
of the sample. It shows that a  continuum ingredient flatter than the mean 
spectrum is the dominant component that characterizes each spectrum. 

The  second component (Fig. 4c) is flat at short UV wavelengths but shows a rising 
continuum redwards of the emission feature at \lam1550\AA~ with a residual 
of P-Cygni profile.
This component represents the slope variation of the SED. It can be associated with 
the UV emission  of the unresolved stellar population, such as individual stars 
and low luminosity stellar clusters (Calzetti, 1999), and it is correlated with the CIV 
\lam 1550 emission.

The main feature of the third component (Fig. 4d) is the rising continuum blueward  
of \lam1400 towards the far-UV region, steeper than  the mean spectrum.
The  modulation by the absorption lines of the  SiIV \lam1400 and of CIV \lam1550 doublets 
 is in the opposite sense to that shown by the PC2.

The rising UV continuum toward the far UV  is recognized as a signature of the 
presence of young massive OB stars. These stars in  which the   emission peaks 
in the far UV,  \lam$\leq1000$\AA~  represent, therefore, 
the bulk of the most recent or current starburst events in the galaxy.
The  strong  winds of these stars originate the P-Cygni profiles seen  in 
SIV and CIV (Mass-Hesse \& Kunth, 1998). 
Therefore, the third component represents the star formation activity and is correlated
with the SIV and CIV absorptions features.

The fractional variance accounted for by the 2nd and 3rd components is similar,  
showing that in our sample the contributions of the two continuum components, 
the young  and the older stellar ones, are similar.

The fourth component (not shown here) is flat and  shows only a weak signature
 of the HeII \lam1640 
absorption/emission. This feature  appears clearly only in the 6th and the 7th principal 
components that contribute only a few per cent (two and one) to the total variance.

\subsection{The normalised sample}

We  applied the normalization by the scalar product  to our sample of spectra
and repeated the PCA analysis. 
The average spectrum of the normalised sample is dominated by the  noise 
(see Fig. 3, tracing b)  and many features are smeared out. In fact, this  normalization
process overweighs the poor signal-to-noise spectra in the sample and the resulting 
average spectrum depends on the  percentage of good signal-to-noise spectra
in the sample. 

The principal components of the normalised sample are shown in Fig 5.
It can be seen that the noise is so high that it hides all the spectral
features. 
The first  component accounts for only 36 per cent of the variance,  
the second one  for  24 per cent, and the third for 13 per cent.
Moreover, the  spectral features of these components are anticorrelated  
%ligth curves
with those of the average spectrum and are similar  to each  other. It seems that  
the PCA technique is unable to  discriminate among the different contributions to 
the SED,  as found for the non-normalised sample.

\begin{figure}
%\begin{minipage}{80mm}
% fig5
\centerline{\epsfxsize=3.5in\epsfbox{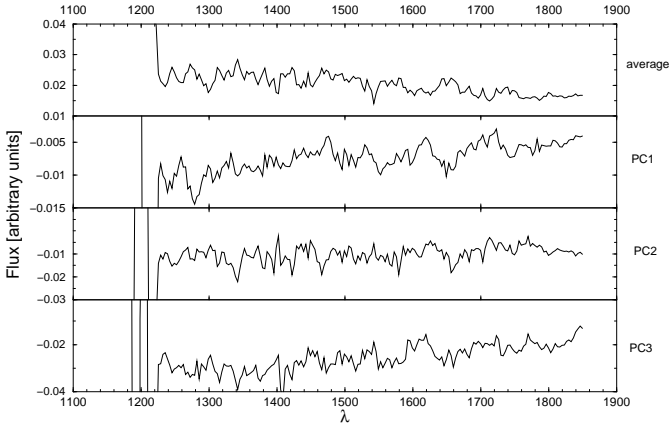}}
\caption{The average spectrum and the principal components for the normalised  set}
%\end{minipage}
\end{figure}

Folkes et al. (1996) found that in a noisy set  of spectra  the PCA method is
still able to select the correlations in the fluxes at each wavelength, but
few  PCA components will be significant.
Our result shows that the very noisy data and the statistical distribution of
different spectral types in our sample make  the normalization technique
unsuitable.

The problem of signal-to-noise limitations of the {\it IUE} data can be  overcome 
by grouping objects with similar spectra into morphological templates 
with high  signal-to-noise.
This binning procedure has been adopted, for instance, by 
Bonatto et al. (1995, 1996, 1998, 1999). 

Instead of grouping the spectra a-priori, we performed a selection by 
signal-to-noise, and applied the PCA method to a sub-sample  of relatively high-quality 
spectra (S/N $\ga$3)  normalised by the scalar product. The average spectrum for this 
sub-sample of 76 galaxies is quite similar to that of the entire 
non-normalised  sample  (Fig. 3, tracing c).
However, the two samples have a different morphological  composition (Fig. 6). 
Selecting only  spectra with good UV signal-to-noise has the  effect of 
increasing the number of irregular/interacting systems, and 
the relative fraction of different galaxy types changes significantly.
The elliptical galaxies are very poorly represented in this sub-sample, since their
number drops from 28 to only 6 galaxies. While the fraction of BCD and compact
galaxies remains almost the same, that of the elliptical ones changes from 24 per cent 
to 10 per cent.

\begin{figure}
%fig 6
\centerline{\epsfxsize=2.in\epsfbox{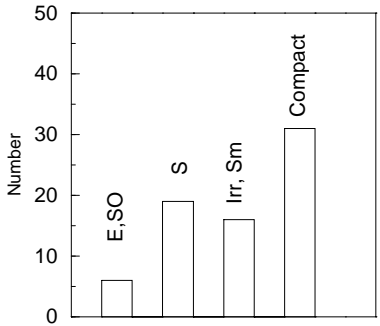}}
\caption{The distribution of spectral types for the good S/N sub-sample}
\end{figure}

The principal components  for this sample are shown in Fig. 7, with
the first component accounting for  41 per cent of the total variance, the 
second one for 23 per cent and the third one for  15 per cent.
The first and, to a lesser degree, the second components show a redwards rising 
continuum that could
represent the old stars' contribution to the total flux, while in PC3 
the continuum is rising bluewards. 
In section 5 we will correlate the PCA results with the
morphological Hubble classification.

\begin{figure}
%\begin{minipage}{80mm}
%fig7
\centerline{\epsfxsize=3.5in\epsfbox{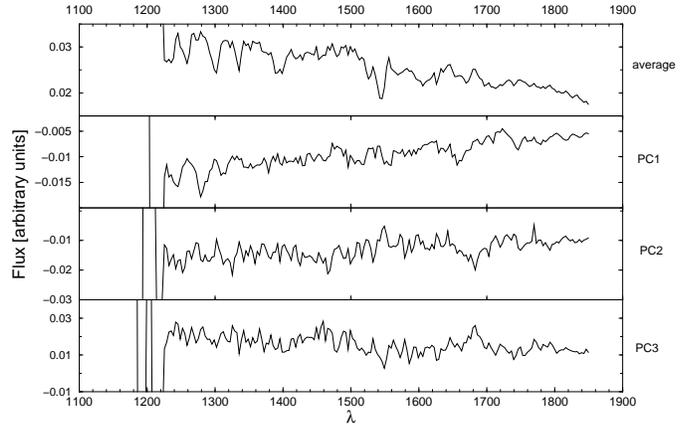}}
\caption{The average spectrum and the principal components for the normalised  subset
of good S/N spectra}
%\end{minipage}
\end{figure}

\subsection{The reduced sub-sample} 

The PCA analysis was performed also for a sample of {\it IUE} galaxies having
the "coverage parameter" (see section 2) up  to 1. For this sample,
the {\it IUE} aperture contained more than  10 percent of the galaxy.
This sample consists of only 43 galaxies, mostly compact, BCD, and EmLS. 
Only four  of the galaxies are normal ellipticals and spirals, and three are 
classified as irregulars. This sample is not 
representative of the galaxy morphology  as a whole, but only of the 
compact and BCD classes.

Fig. 8 shows the principal components  for this sub-sample, where PC1 accounts for 51 
 per cent of the variance, PC2 for 22 per cent, PC3 for 16 per cent 
%and PC4 for 6 per cent 
of the variance, respectively. The first three components allow the reconstruction
of the spectrum with  an accuracy of 89 per cent.

The first  principal component, flatter than the average spectrum, represents the
residual continuum and measures how much each continuum spectrum differs from
the mean one. The second one showing a rising continuum towards long wavelengths
can be the emission of a residual population of A and F stars formed in previous
bursts of star formation. Actually, the  unresolved stellar population of a galaxy
in which bright star-forming clusters, as well as  small low-luminosity
stellar clusters are embedded, contributes with a  large percentage (50-80 per cent) 
of the UV  emission (Calzetti, 1999). 
Note the anticorrelation with the high-ionization absorption lines of SiIV and CIV,  
formed in the expanding atmosphere of massive OB stars, that are a signature of
active star-forming regions. The third component, characterized by a hard UV continuum
and the SiIV and CIV lines in absorption, represents the starburst  activity 
occuring in the galaxy.

\begin{figure}
%all_comp_43.xvgr  fig8
\centerline{\epsfxsize=3.5in\epsfbox{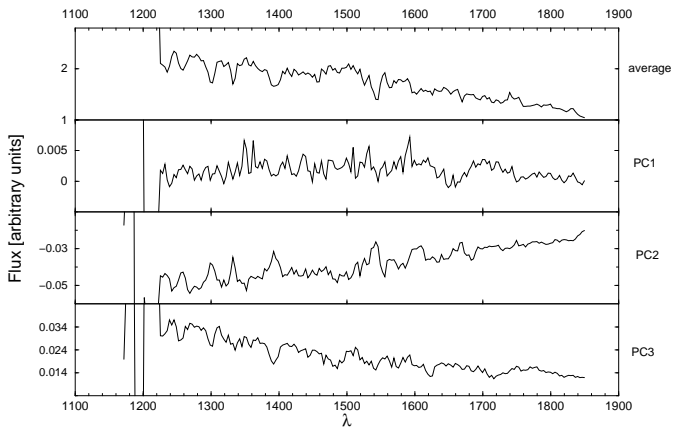}}
\caption{The average spectrum and the principal components for the sample
with coverage factor up to 1}
\end{figure}

\section{ Spectral and morphological classification}
%\section {Discussion}
The PCA anlysis has already been  applied to the study of integrated spectra of
galaxies in the optical range.  Connoly et al. (1995) analyzed a template UV-optical 
spectral dataset that covers the range from 1200\AA~to 1$\mu$m finding that a 
linear combination of two eigenspectra  describes adequately the galaxy spectral types. 
Folkes et al. (1996),  and Sodr\'{e} and Cuevas (1994, 1997), applied
the method to the Kennicutt spectrophotometric atlas of galaxies (Kennicutt, 1992)
and to a sub-sample of normal galaxies. They found that the coefficients  of the 
projection of the spectra onto the principal plane, i.e., the plane defined by the 
first and second principal components, define a spectral sequence.
This sequence  correlates with the Hubble morphological sequence, although there 
is an overlap between the different morphological groups. 

The interpretation of the PCA analysis of the {\it IUE} spectra is  more intriguing.
We analyzed the  coefficients of the projection onto the new orthogonal axes, i.e., 
the PC components, for each spectrum. These coefficients  represent the 
contribution of each  component to the overall spectral distribution of the galaxy.
If there is an underlying global regularity responsabile for the different kinds
of spectra, these coefficients will show correlations and could
be used to define some kind of spectral sequence in the UV.

For the analysis of the  projected spectra, the galaxies have been binned in four 
morphological bins: elliptical, spiral, irregular, and a group containing all the 
compact objects,  namely  galaxies classified  BCD, EmLS, HII,  and compact.
This class includes a variety of objects that show prominent UV emission, which  
can be explained by the presence of a central region of star formation.

It is usual to consider the diagram of the projections on the principal plane,
 since this is the  plane that contains the maximal contribution to  the  variance 
(69  per cent, in our case).  However, the projections on this plane fail to  
reveal any correlation 
and show only a large scatter. The plane defined by the first and  the third 
eigenvectors  represents nearly   the same variance contribution
(66 per cent) and here there is an indication of a sequence.  Note that, as 
described in section 4, the third principal component represents the  contribution 
of young stars to the SED, while the second one is the contribution of the 
underlying background of star forming activity that occurred in the past.

Fig. 9 shows the projections onto the plane defined by the  first and  the third
principal components  for the spectra of the non-normalised sample, after
excluding a few unusual points. 
Although there is an indication of a sequence that follows a direction along the first
component, there is also some vertical  spread along the third  component.
In the plane defined by the second and the third components (fig.10) 
the ellipical galaxies and the compact ones lie on the same region  
apparently tracing  a sequence, while the spiral and irregular ones
show a large scatter.
This could be interpreted as a similarity between the star formation activity of the 
nucleus of the elliptical galaxies and the compact ones.

The SED  of early-type galaxies is often characterized by  rising flux
shortward of $\lambda$1800 \AA, known as the UV turnup. A mixture of
low mass-stars on the horizontal branch evolutionary tracks is 
responsible for the UV flux (e.g., Brown et. al 1997).
This turnup is mostly observed in galactic nuclei and could be signficant
in a sample of nearby galaxies observed troughout the IUE 
aperture, like this one. We cross-checked our sample with 
the Bica et al (1996) sample of elliptical galaxies finding 
only one galaxy in common, NGC4853, which does not present 
a UV turnup. A visual examination of the spectra of 
all the elliptical and SO galaxies of our sample reveals only a few galaxies 
with a moderate UV turnup: NGC1510, ABCG400A and ABCG1795.
Deharveng, Boselli \& Donas (2002), analysed the broad band UV-V and B-V colors for 
a sample of 82 galaxies, finding that only a minority of galaxies shows a strong turnup. 
Note also that  signs of unusual activity have been found in the 
optical spectra of some early-type galaxies (Caldwell et al 1993), 
and residual starburst activity was also claimed by Deharveng, Boselli 
\& Donas (2002)  for some very blue eliptical galaxies.
Although our sample is biased towards the UV brightest galaxies, it is hard
to believe that almost all the elliptical galaxies of the sample show the same level
of nuclear activity as the compact ones.

\begin{figure}
%com1_3_nonorm.xvgr   fig 9
\centerline{\epsfxsize=2.2in\epsfbox{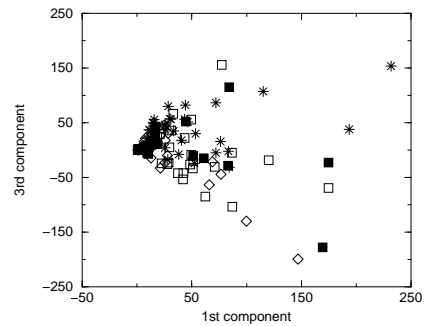}}
\caption{Projections onto the first and third eigenvectors of the spectra 
of the non-normalised sample. Filled squares indicate elliptical 
galaxies, open squares the spiral, diamonds the irregulars, and stars the BCD, 
HII and compact galaxies.}
\end{figure}

\begin{figure}
%com1_3_nonorm.xvgr   fig 10
\centerline{\epsfxsize=2.2in\epsfbox{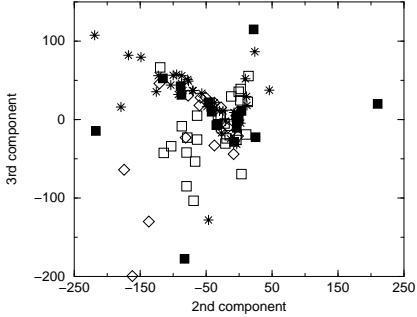}}
\caption{Projections onto the second and third eigenvectors of the spectra 
of the non-normalised sample. Filled squares indicate elliptical 
galaxies, open squares the spiral, diamonds the irregulars, and stars the BCD, 
HII and compact galaxies.}
\end{figure}

Fig. 11 shows the projection onto the plane defined  by the first and the third 
eigenvectors of the spectra of the  sample selected for good signal-to-noise  
and normalised by the scalar product  (section 4.1). 
Here, the  projections lie on an arc sequence,  similar to the spectral sequence
found by Connolly et al. (1995) for the ten templates of starburst and quiescent 
galaxies from Calzetti et al. (1994). Note that the curved 
sequence is due to the scalar 
product normalization, as already pointed out by Sodr\'{e} \& Cuevas (1997).
The sequence is traced by the compact and irregular galaxies, but the binning with
the morphological type is not unique.  Spiral and elliptical galaxies are not segregated
from  other types  and their location in the diagram shows that they are more 
consistent with a significant  starburst component contribution to the SED.

The mixing of spectral types along the  sequence is also evident in the
projection of the components for the  sub-sample of galaxies where
more than 10 per cent of the galaxy  area is included in the {\it IUE} entrance aperture
(see section 4.2). 
In this sample, compact types are the dominant population and many 
of these galaxies  are known to host  one major central starburst region. 
Fig.12a and 12b show the projections of the spectra
onto the plane of the first and the third components and on the plane of the second 
and the third components, respectively. 
In both figures it is evident that the third component is able to  trace
a similar sequence populated by most of the compact  galaxies, while the few
spiral and elliptical galaxies in this sample lie at the bottom of the
oblique trend, in a region of low values
of the third component.

This trend is enhanced in Fig. 13, where the third component is plotted versus 
the average of the first and second orthogonal components.
In this figure the galaxies can be separated into a group which follows a quasi 
linear and tight trend  and a   group with scattered  large positive 
 values on the horizontal axis.
The ultracompact object Pox186 lies at the top of the sequence
and the extremely metal-poor  SBS 0335-052 and Tol 0420-414 lie at its bottom.
The object at the top left-side of this figure is  an HII region
in NGC4449, while the irregular galaxy on the rising trend is IC2458 (Mrk 108), 
classified as BCD by Kinney et al (1993).
The scattering at high  positive values of the horizontal variable  can  
result from the presence of dust. It is well known that the galactic extinction 
is not appropriate for correcting the attenuation by dust inside the galaxy. 
For instance, the point at the extreme right-side of Fig.13 is Tol 0645-376
whose principal components are  always unusual in all the diagrams. 
Terlevich et al. (1991) report a high   H$\alpha$/H$\beta$ ratio of 4.28,
larger than for Case B recombination, indicating that dust is present in
the region of line formation.

It is not clear which  is the physical parameter that defines the sequence.
Pox 186 was believed to be the representative of an extreme population of 
isolated  BCD galaxies at their  first episode of star formation. 
Recently,  an  extended halo feature was detected by Doublier et al. (2000). 
The colors of the outer region of the galaxy are consistent with an underlying evolved 
stellar population of late-K and M stars (Corbin \& Vacca 2002), probably the residual of 
a previous mild episode of star formation.
The UV image  suggests the presence of two
colliding clumps (Corbin \& Vacca 2002), while there is no sign of 
this morphology in the optical image.
Several bright star clusters have been detected at the centre of SBS 0335-052 
(Papaderos et al. 1998) and  also an extensive HI envelope. The observed properties 
make its classification as a young galaxy uncertain (\"{O}stlin \& Kunth 2001).
Similarly, in Tol 0420-414 a diffuse  stellar component underlies the starburst, which
is powered by Wolf-Rayet and O stars (Fricke et al. 2001).

It is becoming evident that not all the BCD galaxies are truly  young objects 
(Kunth \& \"{O}stlin 2000). For instance, the presence of an evolved population, 
formed at a continuous low rate or in a previous episode of star formation, is 
required to explain their broad-band colors (Gil de Paz, Madore \& Perunova 2003).
However, in starbursts the star formation knots dominate the
entire galaxy emission, and only very large telescopes can  detect
halo structures.

In section 4.2 we argued that the third component is associated with the 
hard UV continuum, and 
is representative of the degree of recent star-formation in the galaxy.
The sequence traced in Fig. 13 is similar in shape to the BPT 
 diagram for galaxy line ratios (Baldwin, Phillips \& Terlevich 1981).
In this diagram, the two reddening-insensitive ratios of ~
[OIII]/\Hbeta ~versus ~[NII]6584/\Halpha~ discriminate among 
emission-line galaxies according to the ionization mechanism, 
i.e., ionization due to ultraviolet radiation of  hot stars
or to a power-law continuum(Veilleux \& Osterbrock 1987). 

In the BPT diagram for more than 55.000 galaxies from the Sloan Digital 
Sky Surveys (Kauffmann et al. 2003, Fig. 1) the star-forming galaxies 
lie on a tight sequence. This sequence is correlated  with the 
ionization parameter and the metallicity, for which   the ratio 
[NII]6584/\Halpha~ is a direct estimator (Denicolo' et al 2002).

We checked if the similarity in shape reflects a correlation
of the principal components with the line ratios of the BPT
diagnostic diagram. Spectroscopic data are available for only
very  few galaxies, mainly from Terlevich et al. (1991).
A correlation  with the metallicity can be excluded, while 
no conclusive relation with the [OIII]/\Hbeta ratio  can be  
deduced from so few points.

If the sequence in Fig. 13 is tracing the starbursting
activity, it should correlate with the  luminosity of the Balmer
 \Halpha~line, which is a direct estimator  of the presence of young
massive stars (Kennicutt 1998). 
We found in the literature the  \Halpha~equivalent width (EW) for a few galaxies 
of this sub-sample and we plotted in Fig. 14  the retrieved values 
versus the third component values. 
Although visually there is a hint of a possibile correlation, its statistical
significance is very low. 
More such observations and dust attenuation  correction 
are necessary before  establishing such expected correlation. 
However, the ability of the PCA  method  to separate different
contributions to the SEDs could be used as a  tool for tracing sequences 
in larger samples of starburst galaxies.

\begin{figure}
%comp_1_3_sig.eps   fig 11 
\centerline{\epsfxsize=2.in\epsfbox{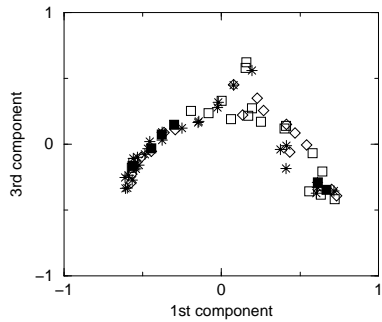}}
\caption{Projections onto the plane defined by the first and the third eigenvectors
of the spectra of the normalised and S/N selected sample. Filled squares indicate elliptical 
galaxies, open squares represent spirals, diamonds  irregulars, and stars BCD, 
HII and compact ones. }
\end{figure}

\begin{figure}
%\begin{minipage}{70mm}
%  fig 12 
\centerline{\epsfxsize=3.8in\epsfbox{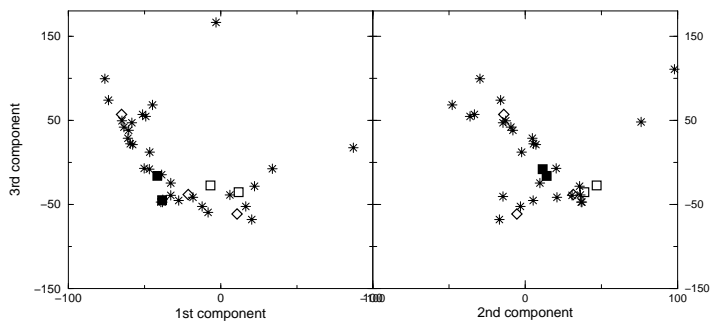}}
\caption{Projections of the spectra of the sub-sample with coverage parameter 
C$\leq1$ onto  a) the plane defined by the first and the third eigenvectors and
 b) the plane defined by the second and the third eigenvectors.  Filled squares 
indicate elliptical galaxies, open squares the spiral, diamonds the irregulars,
 and stars the BCD, HII and compact ones }
%\end{minipage}{80mm}
\end{figure}

\begin{figure}
%  fig 13
\centerline{\epsfxsize=2.2in\epsfbox{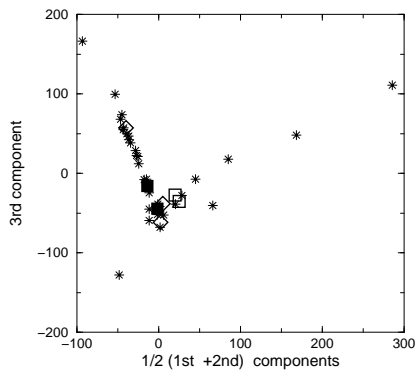}}
\caption{Projections into the plane defined by the linear combination of
the first and the second eigenvectors,  and the third eigenvectors
of the spectra of the sub-sample with coverage parameter up to 1.  Filled squares 
indicate elliptical galaxies, open squares the spiral, diamonds the irregulars and 
stars the BCD, HII and compact ones }
\end{figure}

\begin{figure}
\centerline{\epsfxsize=2.in\epsfbox{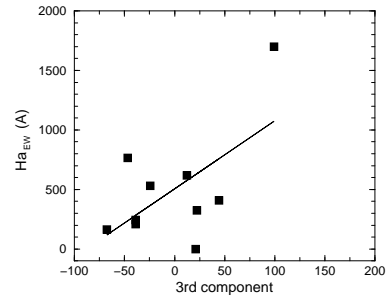}}
\caption{The third component versus \Halpha~for some of the compact galaxies.
The line shows  a linear regression. }
\end{figure}

The data set used was analyzed without any preliminary selection, except for the 
objective coverage factor, defined in section 2. We specifically avoided any attempt 
to  divide the sample into sub-groups according to their spectrum, because in the UV 
some signatures for the presence of young stars are expected for all Hubble types.
It is well-known that episodes of star formation occur in a variety 
of environments, such  as nuclei of spiral galaxies,  or blue compact galaxies, or 
HII galaxies or regions.

We searched for an objective parameter that could define a sequence in the UV
by projecting the spectra into the principal components plane, such as that found
in the optical domain (Connolly et al. 1995; Folkes et al. 1996; Galaz \& de Lapparent 1998).
The UV domain was included only in the analysis of ten templates by Connolly et al. (1995).
However, the UV portion of the SED has a small weight in the spectral range 
of frequencies analysed (from 1200\AA~to 1$\mu$m), and the spectral sequence outlined 
by the PCA  for this sample, is essentially defined by the optical emission of the 
galaxies.

In our data set, the small size of the  entrance aperture of the {\it IUE} could be one of 
the causes of the 
failure of tracing an ordered and continuous sequence of the projections of spectra
into the new axes. 
This aperture sampled  only  the emission from the optical 
centre of the light, while often the brightest part of the UV image
of a galaxy is rather distant  from the centre. 
In nearby spiral galaxies, for instance, the {\it IUE} sampled essentially 
the bulge emission and only partially the  disc emission. Since the circum-nuclear 
bulge hosts a variety of components, including star-forming rings or resolved star 
clusters and spiral structures (Boker et al. 2001), 
it is not surprising that many objects classified as   spiral galaxies  and compact
or BCD galaxies populate the same region in the plane of the  principal components
 (see figs. 9, 10)

\section*{Concluding remarks}
In this paper we analyzed the UV spectral energy distribution of normal
galaxies  listed in  the {\it IUE} INES Guide No. 2-Normal Galaxies 
(Formiggini \& Brosch, 2000) using a Principal Component Analysis.  
Since this method can reveal the internal correlations of a set of data and 
retrieve their common features, we aimed to identify an objective parameter able 
to define a galaxy sequence in the UV.  

The sample consists of the SW-{\it IUE} spectra of the central regions 
of  118 galaxies, covering  a wide range of Hubble types, where the aperture
included more than 1 per cent of the galaxy area.
The PCA method was applied to the sample  without normalisation or scaling, 
to a sub-sample of good S/N spectra, normalised by the 
scalar product, and to a sub-sample where the {\it IUE} aperture contained 
more that 10 per cent of the optical galaxy area.
The principal components retrieved by the PCA have been associated with the main
components observed in the UV spectra of galaxies.
The first component, accounting for the largest source of diversity, can been 
associated with the UV  continuum emission.
The second component, rising redward of 1500\AA, represents the UV contribution 
of an underlying evolved  stellar population (A stars and later). 
The third component is sensitive to
 the amount of activity in the central regions of
galaxies and measures the strength of  recent or current  star formation events.

In all the sample and the sub-samples analysed, the principal component representative
of star-forming activity accounts for a  signficant percentage of the variance.
In the entire sample, the first three principal components explain 83 per cent of the
variance among the spectra.

The  fractional contribution to the SED by the evolved stars and by the young 
population are similar.

The projection of the spectra on the plane defined by the first and the third 
components for  the entire sample fails to outline a regular pattern.
For all the   morphological types, a contribution from  a young population seems
to be present in the nuclear region of the galaxies, although with a different 
activity state.
However none of the principal components is able to outline an internal correlation 
or to correlate with the optical  morphological types.

In a sample of 76  good S/N spectra, normalised by the scalar product, the 
projected spectra show an arc-like sequence, similar to that found by Connolly et
al. (1995) for  the ten template spectra from  the Kinney et al. (1993) sample although 
with a large overlap of morphological classes.

A reduced sample of 43 galaxies, where the {\it IUE} aperture included 
more than 10 per cent of the galaxy, contains almost only  compact and
BCD galaxies. The projections of the spectra of  this sample onto the plane
defined by the first and the third  principal components  outline a sequence
that could be related to  the degree of starburst activity of the galaxy, where
the most active galaxy is Pox 186 and the few early and late-type galaxies 
occupy the low-activity region. A similar sequence is shown by the projections
onto the second and third components.  A quasi-monotonic sequence is outlined 
by the third component with  respect to a  linear combination of the first and second 
components.  The  correlation of the third component  with the \Halpha ~EW 
is not statistically significant, due to the small number of points. We suggest
that more observations could establish that the third principal component 
indeed  represents the star-forming activity of the galaxy.

We initiated this study as an attempt to predict  the UV SED using the
morphological classification of a galaxy and its optical emission properties, using 
the {\it IUE} spectra as the data base.
The work described here indicates that some degree of success in this endeavor 
could be expected, provided that the UV data are based on good signal-to-noise UV
spectra that represent the total UV emision of a galaxy.

Assuming that a 
characterization of the UV properties would be done from the {\it GALEX} results, the 
galaxy coverage would be taken care of automatically by using the objective prism
feature. However, one would have to deal with
low-resolution spectra  even for point-like objetcs ($\simeq10$\AA~at CIV \lam1550, 
more than twice that used
here) and it is not clear how this would influence the results. Given the lack of
spectral discriminators longwards of $\simeq2000$\AA, it is unlikely that 
the LW part of the  {\it GALEX} spectra would add much information content to 
a future PCA analysis.

\begin{table*}
\caption{Table 1- List of galaxies   } 
\begin{tabular}{@{}lllllllc@{}} \\
\hline
Id.no.&  & $\alpha$$_{2000}$ &$\delta$$_{2000}$ & Coverage & B$_{T}$ &  z & Morph. Type \\  
\\
\hline
 NGC7828                & 1   & 00 06 27& -13 24 54&  1.45 &   14.37  & 0.01935  &  Sc              \\ 
  NGC118                & 1   & 00 27 16& -01 46 48&  1.34 &   14.63  & 0.03753  &  I0              \\ 
  ESOB350-IG38          & 1   & 00 36 53& -33 33 14&  0.00 &   15.45  & 0.02053  &  S?              \\ 
  ABCG85                &     & 00 41 51& -09 18 15&  1.92 &   14.50  & 0.05567  &  cD;SB0          \\ 
  ESOB474- 26           &     & 00 47 08& -24 22 14&  1.78 &   14.94  & 0.05271  &  SA              \\ 
  IC1586                & 1,2 & 00 47 56& +22 22 28&  0.00 &   14.90  & 0.01942  & Compact          \\ 
  AOO PKS0123-16        & 1,2 & 01 25 48& +01 22 18&  0.00 &   14.42  & 0.01875  &  I               \\ 
  MRK2                  & 1   & 01 54 53& +36 55 02&  1.42 &   13.92  & 0.01875  &  SB0a            \\ 
  NGC835                & 1   & 02 09 25& -10 08 09&  1.91 &   12.97  & 0.01359  &  SAB             \\ 
  IC214                 & 1   & 02 14 06& +05 10 32&  0.00 &   14.68  & 0.03022  &  S               \\ 
  NGC992                & 1   & 02 37 26& +21 05 55&  1.60 &   15.56  & 0.01381  &  S?              \\ 
  MRK600                & 1,2 & 02 51 04& +04 27 09&  0.70 &   14.99  & 0.00336  &  SBb;BCD         \\ 
  ABCG400A              &     & 02 57 42& +06 01 38&  1.20 &   13.86  & 0.02215  &  E               \\ 
  AOO SBSG 0335-052     & 1,2 & 03 37 47& -05 02 47&  0.46 &          & 0.01349  & EmLS             \\ 
  ESOB156-IG7           &     & 03 41 12& -54 00 39&  1.06 &   15.71  & 0.05306  &  G               \\ 
  NGC1510               & 1   & 04 03 33& -43 24 01&  1.76 &   13.51  & 0.00304  & SA0              \\ 
  AOO TOL 0420-414      & 1,2 & 04 21 59& -41 19 21&  0.00 &   18.26  & 0.01993  &  EmLS            \\ 
  AOO TOL 0440-381      & 1,2 & 04 42 08& -38 01 03&  0.00 &   18.26  & 0.04100  & EmLS             \\ 
  MRK1094               & 1   & 05 10 48& -02 40 54&  1.42 &   13.87  & 0.00944  &  I0 pec?         \\ 
  AOO TOL 0513-393      & 2   & 05 15 20& -39 17 41&  0.00 &   17.41  & 0.05000  &EmLS              \\ 
  AOO 0644-741          &     & 06 43 00& -74 14 11&  1.93 &   13.88  & 0.02170  &  E               \\ 
  AOO TOL 0645-376      & 2   & 06 46 49& -37 43 25&  0.00 &   17.38  & 0.02600  &  EmLS            \\ 
  MRK7                  & 1   & 07 28 11& +72 34 20&  1.31 &   14.44  & 0.01021  &  S               \\ 
  IC2184                & 1   & 07 29 25& +72 07 41&  0.00 &   14.38  & 0.01202  &  S               \\ 
  AOO HARO 1            & 1   & 07 36 57& +35 14 33&  1.71 &   12.72  & 0.01262  &  Im?             \\ 
  MRK12                 & 1   & 07 50 48& +74 21 32&  1.80 &   13.11  & 0.01318  &  SAB             \\ 
  PG 0833+652           & 1,2 & 08 38 23& +65 07 16&  0.00 &          & 0.01911  &  pec  HII        \\ 
  NGC2623               &     & 08 38 24& +25 45 01&  0.00 &   13.99  & 0.01846  &  S               \\ 
  AOO T 0840+120        & 1,2 & 08 42 21& +11 50 01&  0.00 &          & 0.03000  &  G               \\ 
  MRK702                & 1   & 08 45 34& +16 05 48&  1.34 &   15.57  & 0.05280  & Compact          \\ 
  NGC2684               &     & 08 54 53& +49 09 38&  1.66 &   13.65  & 0.00954  &  S?              \\ 
  NGC2773               & 1   & 09 09 44& +07 10 26&  1.12 &   14.46  & 0.01834  &   S?             \\ 
  IC2458                &1,2  & 09 21 29& +64 14 11&  0.80 &   15.41  & 0.00512  &  I0 pec          \\ 
  MRK116 A              &1,2  & 09 34 02& +55 14 25&  0.00 &   15.61  & 0.00248  &  Compact         \\ 
  ESOB435-IG20          & 1   & 09 59 21& -28 07 54&  1.48 &   14.40  & 0.00237  &  Sb              \\ 
  MRK25                 & 1   & 10 03 52& +59 26 11&  1.20 &   14.82  & 0.00868  &  Sb?             \\ 
  NGC3125               & 1   & 10 06 34& -29 56 10&  1.69 &   13.47  & 0.00288  &  S:BCDG          \\ 
  MRK26                 &1,2  & 10 11 51& +58 53 31&  0.58 &   15.99  & 0.03043  &  Sc              \\ 
  MRK33                 & 1   & 10 32 31& +54 23 56&  1.75 &   12.98  & 0.00487  &  Im              \\ 
  NGC3353               & 1   & 10 45 23& +55 57 33&  1.91 &   13.22  & 0.00315  &  BCD             \\ 
  MRK153                & 1   & 10 49 05& +52 19 58&  1.40 &   14.98  & 0.00805  &  Scp             \\ 
  MRK1267               & 1   & 10 53 04& +04 37 43&  1.20 &   14.16  & 0.01932  &  E?              \\ 
  ABCG1126              &     & 10 53 50& +16 51 00&  0.00 &   15.12  & 0.08584  &  S               \\ 
  NGC3471               &     & 10 59 09& +61 31 51&  1.93 &   13.24  & 0.00710  &  Sa              \\ 
  AOO APG 148           &     & 11 03 54& +40 51 00&  0.00 &          & 0.03452  &   G              \\ 
  MRK36                 &1,2  & 11 04 58& +29 08 22&  0.58 &   15.64  & 0.00215  &  BCD             \\ 
\end{tabular}
\end{table*}
\newpage
\begin{table*}
\begin{tabular}{@{}lllllllc@{}} \\ 
\hline
Id.no.&  & $\alpha$$_{2000}$ &$\delta$$_{2000}$ & Coverage & B$_{T}$ &  z & Morph. Type \\  
\hline
  AOO CASG 816 W        & 2   & 11 12 08& +35 52 43&  0.40 &   17.50  & 0.02844  & Compact          \\ 
  MRK170                &     & 11 26 50& +64 08 16&  1.50 &   15.03  & 0.00329  &  Irr             \\ 
  MCG +13-08-0058       & 1   & 11 28 01& +78 59 29&  1.85 &   15.08  &-0.00033  &  Pec             \\ 
  MRK178                & 1   & 11 33 29& +49 14 12&  0.58 &   14.45  & 0.00083  &   Irr            \\ 
  AOO ARP 248B          & 1   & 11 46 45& -03 50 54&  1.68 &   15.15  & 0.01724  &  SBb             \\ 
  NGC3991               & 1   & 11 57 31& +32 20 00&  1.18 &   13.52  & 0.01105  &  Sc              \\ 
  NGC3994               & 1   & 11 57 36& +32 16 44&  1.58 &   13.32  & 0.01040  &  SA              \\ 
  NGC4004               & 1   & 11 58 05& +27 52 38&  1.83 &   13.97  & 0.01126  &  Irr             \\ 
  AOO POX 36            & 1   & 11 58 59& -19 01 36&  1.50 &   14.28  & 0.00372  &  IBm             \\ 
  AOO HE 1203-2644      & 1   & 12 05 59& -27 00 54&  1.45 &   14.98  & 0.00590  &  HII             \\ 
  IC3017                & 2   & 12 09 25& +13 34 25&  0.88 &   14.82  & 0.00658  &  BCD             \\ 
  AOO VCC 22            & 2   & 12 10 24& +13 1o 24& -0.20 &   16.13  & 0.00564  &  BCD?            \\ 
  AOO VCC 24            & 1,2 & 12 10 36& +11 45 37&  0.80 &   15.10  & 0.00430  &  BCD             \\ 
  MCG +01-31-030        & 1,2 & 12 15 19& +05 45 42&  0.70 &   14.99  & 0.00674  &  E0 pec?         \\ 
  AOO 1214-277          & 2   & 12 17 21& -28 02 32& -0.39 &          & 0.02600  &  EmLS            \\ 
  AOO 1214-28           & 2   & 12 17 17& -28 02 33& -0.39 &          & 0.02600  &  EmLS            \\ 
  MRK49                 &     & 12 19 10& +03 51 28&  1.42 &   14.29  & 0.00509  &  E pec: H        \\ 
  AOO VCC 562           & 2   & 12 22 36& +12 09 28&  0.75 &   16.43  & 0.00015  &  BCD             \\ 
  MRK209                & 1   & 12 26 16& +48 29 31&  1.48 &   14.65  & 0.00094  &  Sm              \\ 
  NGC4449$_{3}$         & 1,2 & 12 28 11& +44 05 40&  3.24 &    9.84  & 0.00069  &  IBm  HII        \\ 
  NGC4449$_{4}$         & 1,2 & 12 28 11& +44 05 40&  3.24 &    9.84  & 0.00069  &  IBm  HII        \\ 
  NGC4449$_{5}$         & 1,2 & 12 28 11& +44 05 40&  3.24 &    9.84  & 0.00069  &  IBm  HII        \\ 
  NGC4449$_{6}$         & 1,2 & 12 28 11& +44 05 40&  3.24 &    9.84  & 0.00069  &  IBm  HII         \\ 
  MGC UGC 7905          & 1   & 12 43 48& +54 53 45&  1.58 &   14.06  & 0.01626  &  S? pec          \\ 
  AOO HARO 33           & 1   & 12 44 38& +28 28 19&  1.25 &   12.72  & 0.00316  &  S0 pec          \\ 
  NGC4650A              &     & 12 44 50& -40 42 54&  1.91 &   13.92  & 0.00954  &  S0              \\ 
  NGC4670               & 1   & 12 45 17& +27 07 34&  1.99 &   13.07  & 0.00357  &  SB0             \\ 
  MCG +02-33-0012       & 2   & 12 46 05& +08 28 31&  1.00 &   14.78  & 0.00495  &  BCD             \\ 
  AOO TOL 1247-232      & 1,2 & 12 50 19& -23 33 57&  0.00 &          & 0.04800  &  EmLS            \\ 
  NGC4774               & 1   & 12 53 07& +36 49 07&  0.00 &   14.81  & 0.02793  &  S               \\ 
  MRK54                 & 1   & 12 56 56& +32 26 55&  1.25 &   15.29  & 0.04477  &  Sc?             \\ 
  NGC4853               &     & 12 58 35& +27 35 50&  1.61 &   14.40  & 0.02555  &  SA0             \\ 
  QSO 1300+361          & 2   & 13 03 03& +35 51 29&  0.00 &   18.00  & 0.06055  &  HII             \\ 
  AOO POX 120           & 2   & 13 06 42& -12 04 22&  0.00 &   15.70  & 0.02075  &  EmLS            \\ 
  AOO POX 124           & 2   & 13 07 26& -13 11 01&  0.00 &   15.51  & 0.02422  &  EmLS            \\ 
  MCG +07-27-0052       & 1   & 13 14 10& +39 08 51&  0.00 &   15.58  & 0.00388  &  GPair           \\ 
  MRK450                &     & 13 14 48& +34 52 44&  1.70 &   14.33  & 0.00285  &  Im?             \\ 
  NGC5122               &     & 13 24 15& -10 39 16&  1.18 &   14.10  & 0.00980  &  Sc              \\ 
  AOO POX 186           & 1,2 & 13 25 51& -11 37 35&  0.00 &   17.00  & 0.00390  &  EmLS            \\ 
  MRK66                 & 1,2 & 13 25 54& +57 15 05&  0.88 &   15.05  & 0.02176  &  BCG             \\ 
  NGC5236$_{1}$         & 1,2 & 13 37 00& -29 52 04&  3.97 &    7.92  & 0.00172  &  SAB  HII        \\ 
  NGC5236$_{3}$         & 1,2 & 13 37 00& -29 52 04&  3.97 &    7.92  & 0.00172  &  SAB  HII        \\ 
  ESOB383-G44           & 1   & 13 37 27& -33 00 22&  1.79 &   14.05  & 0.01260  &  SAd             \\ 
  MRK67                 & 2   & 13 41 56& +30 31 11&  0.40 &   16.36  & 0.00320  &  BCD             \\ 
\end{tabular}
\end{table*}
\newpage
\begin{table*}

\begin{tabular}{@{}lllllllc@{}} \\
\hline
Id.no.&  & $\alpha$$_{2000}$ &$\delta$$_{2000}$ & Coverage & B$_{T}$ &  z & Morph. Type \\  
\\
\hline
  NGC5291               &     & 13 47 24& -30 24 27&  1.69 &   13.64  & 0.01463  &  E pec:          \\ 
  ABCG1795              &     & 13 48 52& +26 35 35&  1.80 &   15.20  & 0.06326  &  cD;S0?          \\ 
  AOO Z 13502+0022      & 1   & 13 52 44& +00 07 51&  1.06 &   15.43  & 0.01209  &  Sm              \\ 
  MCG +04-33-038        & 1   & 14 01 09& +21 14 15&  1.18 &   15.04  &          &  GPair           \\ 
  AOO TOL 41            & 2   & 14 02 59& -30 14 25& -0.30 &   18.00  & 0.02308  &  EmLS            \\ 
  IC4448                &     & 14 40 28& -78 48 37&  1.70 &   14.06  & 0.01543  &  SBd             \\ 
  MRK288                & 2   & 14 50 47& +73 49 24&  0.00 &   15.70  & 0.02502  &  S?              \\ 
  MCG +06-33-00 0004    & 1   & 14 50 57& +35 34 17&  1.75 &  14.77   &0.00405   &  BCDG            \\ 
  ABCG1991              &     & 14 54 32& +18 38 24&  1.83 &   15.29  & 0.05921  &  E               \\ 
  AOO CASEG 657         & 1,2 & 15 12 13& +47 16 31&  0.00 &   16.30  & 0.05350  &  Compact         \\ 
  MRK487                & 1,2 & 15 37 04& +55 15 47&  0.58 &   15.46  & 0.00002  &  Compact         \\ 
  NGC5996               & 1   & 15 46 59& +17 53 08&  1.99 &   13.01  & 0.01100  &  SBc             \\ 
  NGC6090               & 1   & 16 11 40& +52 27 21&  0.75 &   14.49  & 0.02930  &  GPair           \\ 
  MRK499                & 1,2 & 16 48 24& +48 42 23&  0.40 &   14.60  & 0.02567  &  Im:             \\ 
  AOO FRL 44            &     & 18 13 39& -57 43 58&  1.10 &          & 0.01650  &  S? pec          \\ 
  ESOB338-IG4           & 1   & 19 27 58& -41 34 28&  1.31 &   13.42  & 0.00959  &  pec  HII        \\ 
  ESOB185-IG13          & 1,2 & 19 45 01& -54 15 03&  0.88 &   15.00  & 0.01868  &  Compact:HII     \\ 
  ESOB462-IG20          & 1   & 20 26 57& -29 07 06&  0.00 &   14.60  & 0.02011  &  E               \\ 
  ESOB400-G4            & 1   & 20 37 42& -35 29 11&  1.10 &   14.28  & 0.01968  &  Compact:HII     \\ 
  NGC7173               &     & 22 02 04& -31 58 25&  1.83 &   13.14  & 0.00833  &  E pec:         \\ 
  NGC7176               &     & 22 02 09& -31 59 25&  1.70 &   12.42  & 0.00838  &  E pec           \\ 
  NGC7250               & 1   & 22 18 18& +40 33 45&  1.93 &   13.20  & 0.00389  &  Sdm?            \\ 
  MCG -07-47-023        &     & 23 13 59& -42 43 39&  1.82 &   15.51  & 0.05640  &  SA0             \\ 
  NGC7609 B             &     & 23 19 31& +09 30 10&  1.27 &   16.00  & 0.03857  &  Sm              \\ 
  ABCG2597              &     & 23 25 20& -12 07 27&  1.40 &   16.32  & 0.08220  &Elliptical        \\ 
  NGC7673               & 1   & 23 27 42& +23 35 24&  1.99 &   13.10  & 0.01137  &  SAc? pec        \\ 
  AOO Z 2327.6+251      & 1   & 23 30 09& +25 31 43&  0.69 &   15.06  & 0.01923  & Sb               \\ 
  ABCG2626              & 2   & 23 36 30& +21 08 48&  0.70 &   15.31  & 0.05499  &  S0              \\ 
  ABCG2634              &     & 23 38 30& +27 01 51&  0.00 &   13.47  & 0.03022  &  E               \\ 

\end{tabular}
\end{table*}

\section*{Acknowledgments}

We acknowledge helpful discussion with O. Lahav and S. Steindling. UV astronomy at Tel Aviv University is
supported by the Austrian Friends of the Tel Aviv University.

\label{lastpage}

\end{document}